\documentclass[prx,twocolumn]{revtex4}

\usepackage[utf8]{inputenc}
\usepackage{amsmath}
\usepackage{amsthm}
\usepackage{amsfonts}
\usepackage{amssymb}
\usepackage{url}
\usepackage{graphicx}
\usepackage[colorlinks]{hyperref}
\hypersetup{citecolor=blue,
urlcolor=blue}
\newtheorem{theorem}{Theorem}

\newtheorem{definition}[theorem]{Definition}

\newcommand{\ket}[1]{\lvert #1 \rangle}
\newcommand{\bra}[1]{\langle #1 \rvert}

\usepackage{color}

\definecolor{nblue}{rgb}{0.3,0.3,1.0}
\definecolor{ngreen}{rgb}{0.2,0.7,0.2}
\definecolor{nred}{rgb}{0.9,0.1,0}
\definecolor{red2}{rgb}{0.6,0.2,0.2}
\definecolor{npurple}{rgb}{0.8,0.2,0.8}
\definecolor{golden}{rgb}{0.8,0.6,0.1}
\definecolor{nsilver}{rgb}{0.3,0.4,0.5}
\definecolor{nbrown}{rgb}{0.8,0.4,0.15}
\definecolor{nrose}{rgb}{0.7,0,0.35}
\definecolor{nviol}{rgb}{0.5,0,1.0}
\definecolor{nazur}{rgb}{0,0.35,0.7}
\definecolor{nchart}{rgb}{0.2,0.4,0}
\definecolor{nbrick}{rgb}{0.55,0.25,0.15}

\begin{document}
\date{\today}
\title{A possibilistic no-go theorem on the Wigner's friend paradox}
\author{Marwan Haddara}
\affiliation{Centre for Quantum Dynamics, Griffith University, Gold Coast, Queensland 4222, Australia}
\author{Eric G. Cavalcanti}

\affiliation{Centre for Quantum Dynamics, Griffith University, Gold Coast, Queensland 4222, Australia}
\begin{abstract}
    In a recent work, Bong \emph{et al.} [Nature Physics 16, 1199 (2020)] proved a no-go theorem demonstrating a contradiction between a set of assumptions called ``Local Friendliness'' (LF) and certain quantum phenomena on an extended version of the ``Wigner's friend'' paradox. The LF assumptions can be understood as the conjunction of two independent assumptions: Absoluteness of Observed Events (AOE) requires that events observed by any observer have absolute, rather than relative, values; Local Agency (LA) encodes the assumption that an intervention cannot influence events outside its future light cone. The proof of the LF no-go theorem, however, implicitly assumes the validity of standard probability theory. Here we present a probability-free version of the Local Friendliness theorem, building upon Hardy's no-go theorem for local hidden variables. The argument is phrased in the language of possibilities, which we make formal by using a modal logical approach. It relies on a weaker version of Local Agency, which we call ``Possibilistic Local Agency'': the assumption that an intervention cannot influence the \emph{possibilities} of events outside its future light cone.
\end{abstract}
\maketitle

\section{Introduction}
 Bell's theorem~\cite{Bell1964} shows that there exist quantum correlations that cannot be reproduced by any Local Hidden Variable (LHV) model. The correlations demonstrating that impossibility are identified via the violation of Bell inequalities---constraints of the space of possible device-independent correlations that bound the set of LHV correlations. A LHV model can be derived from various alternative sets of premises regarding physical theories~\cite{Wiseman2017, Cavalcanti2021}, and the violations of Bell inequalities can thus be taken to imply the impossibility to construct a theory simultaneously satisfying all premises in any such ``LHV-complete'' set. 
 
The derivation of the Bell inequalities, however, also assumes the validity of standard probability theory to make various inferences from sets of metaphysical premises to the existence of a LHV model, and from a LHV model to Bell inequalities. This opens the question~\cite{Pitowsky1982} whether Bell's theorem may be resolved by rejecting the applicability of  probability theory in those derivations, rather than any of the metaphysical premises.
  
The famous no-go theorems by Greenberger, Horne and Zeilinger (GHZ) \cite{Greenberger1989} and Hardy \cite{Hardy1992, Hardy1993}, among others, show that it is possible to prove a version of Bell's theorem without direct use of probability theory. Instead, those proofs demonstrate the failure of a deterministic local hidden variable model, via a set of logical contradictions involving whether or not it is \emph{possible} for certain events to occur, according to a given theory, rather than their \emph{probabilities}. These have been referred to as proofs of Bell's theorem without inequalities \cite{Greenberger1990}. 
  
Along a recent line of research \cite{Brukner2017, Frauchiger2018, Brukner2018, Baumann2018, Healey2018, AllardGuerin2021} surrounding versions of the `Wigner's friend' thought experiment \cite{Wigner1961}, a new quantum no-go theorem, the Local Friendliness (LF) no-go theorem,  has been introduced \cite{Bong2020}. The LF no-go theorem is similar in spirit to Bell's theorem in the sense that the metaphysical premises going into it are phrased in a manner independent of the details of any particular physical theory or the degrees of freedom of the systems that make up the scenario. The premises underlying the LF no-go theorem are however \emph{weaker} than those of Bell's theorem, and so violations of LF inequalities put strictly stronger constraints on physical theories~\cite{Bong2020,Cavalcanti2021}.

Just as with Bell's theorem, however, the LF no-go theorem uses probability theory to make inferences. Most importantly, it involves assigning a joint probability distribution to sets of variables that can be inferred to have simultaneously well-defined values, given the premise of \emph{Absoluteness of Observed Events}~\cite{Bong2020}. Thus one may wonder whether it could be possible to formulate a probability-free version of the LF no-go theorem, analogous to the Hardy and GHZ theorems.

In this work it is shown that an inequality-free version of the LF no-go theorem can indeed be formulated.  The proof is inspired by observations \cite{Brukner2018,Pusey2018, Aaronson2018} regarding the similarities of Hardy's paradox \cite{Hardy1992, Hardy1993} and the no-go theorem of  Frauchiger and Renner (FR) \cite{Frauchiger2018}, where a logical contradiction arises from certain premises in an extended Wigner's friend scenario. However, some of the assumptions in the FR proof directly refer to quantum theory. In contrast, our result is  theory-independent.

In Ref.~\cite{Nurgalieva2019} the FR no-go theorem \cite{Frauchiger2018} was phrased in terms of \emph{epistemic modal logic} --- a logic designed to deal with statements about knowledge. Their conclusion was that such a  logic is in general inadequate for modelling the knowledge of agents in an extended Wigner's friend scenario. Inspired by that approach, we also formalise our argument using modal logic. However, in this paper the so-called \emph{alethic modalities} \cite{Rantala2004}, namely those of possibility and necessity are considered (here those modalities refer to \emph{physical}, rather than logical or metaphysical, possibility). From this perspective, our approach bears similarities with recent works on inequality-free non-classicality notions such as \cite{Mansfield2012, Abramsky2013, Mansfield2017, Santos2021}, where probability distributions are replaced with ``possibility distributions'', which indicate which outcomes are physically possible and which outcomes are not physically possible in a given experiment. 

\section{The Extended Wigner's Friend Scenario \label{EWFS}} The Wigner's friend thought experiment \cite{Wigner1961} is one of the sharpest ways to illustrate the quantum measurement problem.  This thought experiment considers an observer (the ``friend'')  measuring a quantum system, and a superobserver (``Wigner'') who is describing both the system and the friend according to the laws of quantum mechanics.  Suppose, for example, that the friend performs a measurement on a qubit system $S$. Let system $F$ (which we will henceforth refer to as the Friend) represent the friend and their ``laboratory''---i.e. any degree of freedom the friend interacts with during the experiment, other than the system $S$.

Suppose the Friend performs an ideal projective measurement of $S$ in the computational basis $\ket{0}_S$, $\ket{1}_S$. Let $\ket{R}_F$ be a state vector of system $F$ in an initial ``ready" state for such a measurement. Then if $S$ is initially in state $\ket{0}_S$, the state of the Friend will change to $\ket{0}_F$, representing the Friend having observed outcome `0'. If $S$ is initially in state $\ket{1}_S$ $F$ will instead change to $\ket{1}_F$.  

If instead $S$ is initially prepared in a superposition $\alpha \ket{0}_S + \beta \ket{1}_S$, then according to the ``collapse postulate'' the joint state of $F$ and $S$ undergoes a stochastic map of the kind 
\begin{align} \label{collapsestate}
    (\alpha \ket{0}_S + \beta \ket{1}_S) \otimes \ket{R}_F \xrightarrow{\text{C}} 
            \begin{cases}
                \ket{0}_S \ket{0}_F \;, \; p=|\alpha|^2\\ 
                \ket{1}_S \ket{1}_F \;, \; p = |\beta|^2\,.
            \end{cases}
\end{align}

On the other hand if the universal validity of quantum mechanics is assumed, and the joint system $S+F$ is isolated, the evolution due to the measurement by the Friend ought to be unitary, leading to 
\begin{align}\label{unitarystate}
    (\alpha \ket{0}_S + \beta \ket{1}_S) \otimes \ket{R}_F \xrightarrow{\text{U}} \alpha \ket{0}_S \ket{0}_F + \beta \ket{1}_S \ket{1}_F.
\end{align} 
Wigner concluded~\cite{Wigner1961} that ``this is a contradiction, because the wave function \eqref{unitarystate} describes a state that has properties which neither [of the states in \eqref{collapsestate}] has''.

The original Wigner's friend scenario as well as some modified versions of it are usually discussed in fully quantum mechanical contexts. An abstract, theory independent picture of the experiment can however, lead to very interesting and strong no-go results, as was shown, for example in Ref.~\cite{Bong2020} by building on a version of the scenario introduced  by Brukner \cite{Brukner2017, Brukner2018}.

\begin{figure}
    \centering
    \includegraphics[scale=0.2]{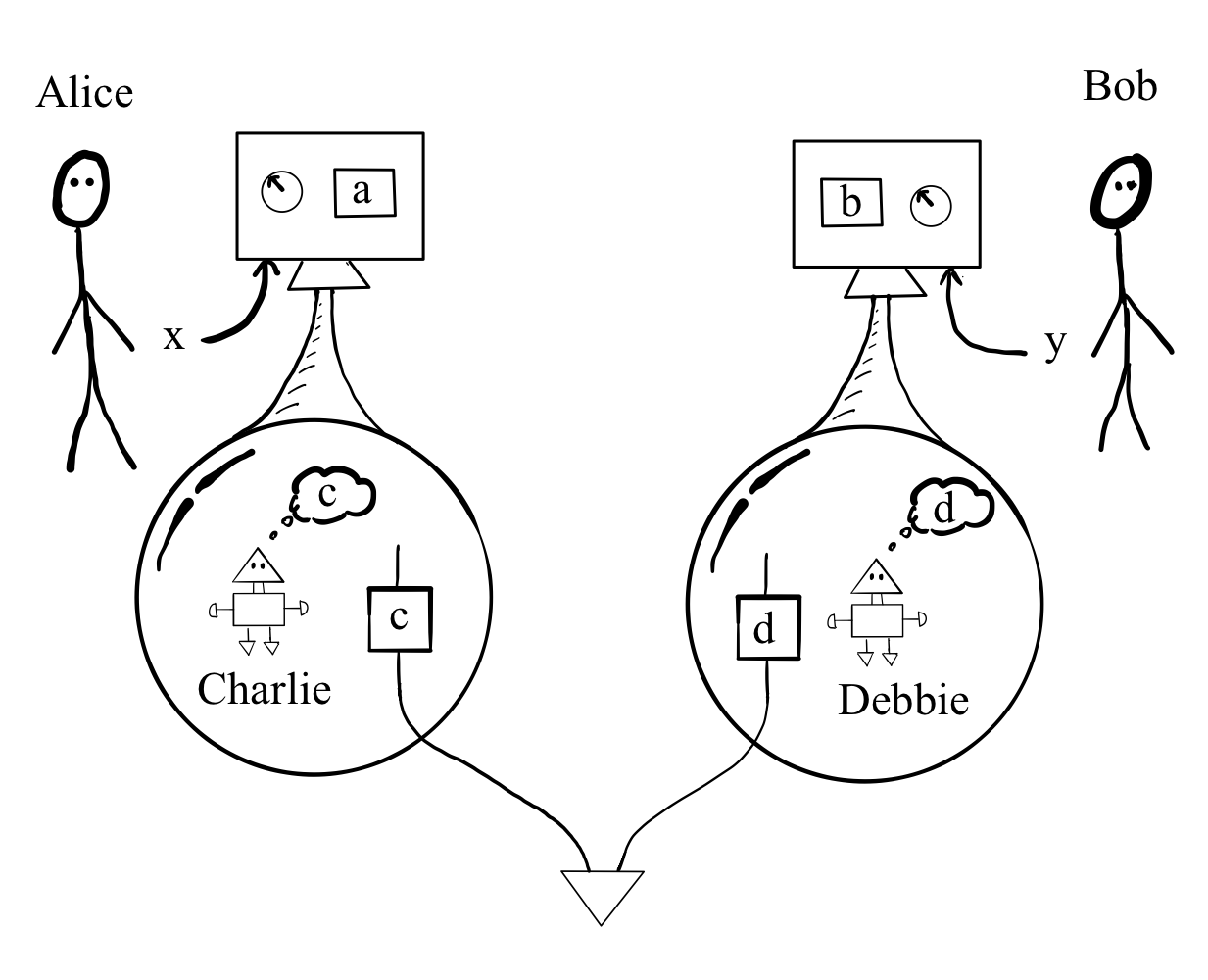}
    \caption{Observers Charlie and Debbie record outcomes $c,d$ in their completely isolated laboratories, depicted here as bubbles. Superobservers Alice and Bob can perform arbitrary measurements labelled $x,y$ on Charlie and Debbie obtaining outcomes $a,b$ respectively. This figure is reproduced with permission from \cite{Cavalcanti2021Foundphys}. }
    \label{fig:2_LF_setup}
\end{figure}

For the elucidation of these extended scenarios, it is assumed that the experiments can be described in a relativistic space-time, specifying a light-cone structure. An abstract ``general'' EWFS can be taken  to consist of a number of space-like separated superobservers, some of which have ``friends''. Those superobservers that do not have a friend can be thought to measure some system directly. An ordinary Bell-scenario may then be interpreted as a special case of an EWFS, in which none of the superobservers have friends. One measurement for the superobservers who have a friend will be given a special role; it can be thought to represent the situation where the superobserver asks for the record of the measurement made by his friend and assigns that value to their own measurement outcome. The other measurements can be regarded as arbitrary.  Different theories might give different accounts of what is happening in an instance of such an experiment, but any empirically adequate theory should be able to reproduce the predictions for the outcomes of measurements made by the superobservers, which constitute the empirical data in such an experiment.

\begin{figure}
    \centering
    \includegraphics[scale=0.38]{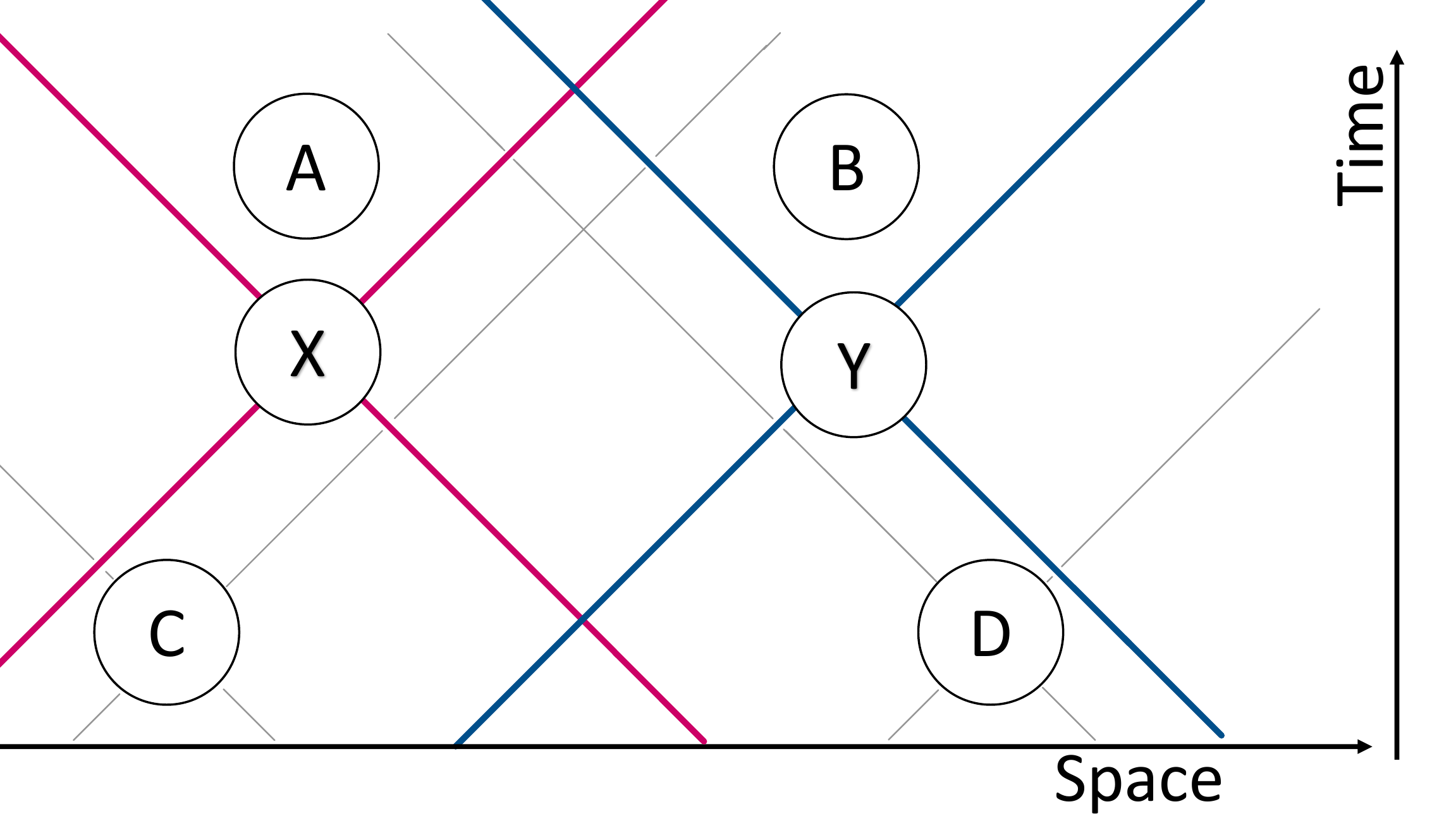}
    \caption{Spacetime diagram for the type of two-party two-friend EWFS considered in this paper. $C,D$ correspond to observations made by Charlie and Debbie. Observation of the variable $A$ is in the future light cone (FLC) of the intervention $X$ by Alice, resp. $B$ is in the FLC of $Y$ by Bob. Alternatively $X$ is not in the past light cone (PLC) of $B,Y,C,D$  and $Y$ is not in the PLC of $A,X,C,D$. }
    \label{fig:spacetime}
\end{figure}

In this paper we consider a particular instance of such an EWFS where two superobservers,  Alice and Bob, are performing experiments at space-like separation. Alice and Bob each have a nearby friend, Charlie and Debbie respectively.  Let $X,Y$ denote the intervention variables for Alice and Bob with values $x, y \in \{1,2 \}$. Let $A,B$ denote the observations of Alice and Bob with values  $a,b \in \{0,1 \}$ for each measurement. The intervention $X=1$ corresponds to Alice opening the door to Charlie's lab, which is otherwise sealed shut, and asking for the result $C=c \in \{0,1 \}$ and then assigning $A = c$. Similarly the setting $Y = 1$ of Bob corresponds to asking Debbie for her result $D=d \in \{0,1 \}$. The setup is portrayed in Fig. \ref{fig:2_LF_setup}    while the spatio-temporal relations of the type of scenario described here are illustrated in Fig. \ref{fig:spacetime}.

\section{The Local Friendliness model}``Local Friendliness'' can be stated as a conjunction of two principles: Absoluteness of Observed Events and Local Agency \cite{Bong2020, Cavalcanti2021}.

\begin{definition}[Absoluteness of Observed Events (AOE)]
Every observed event is an absolute single event, not relative to anything or anyone. \label{AOE}
\end{definition}

\begin{definition}[Local Agency (LA)]
Any intervention is uncorrelated with any set of relevant events outside its future light cone. \label{LA}
\end{definition}

The term `intervention' in Def.~\ref{LA} refers to a choice of values for variables that are assumed to be controllable by experimenters. An intervention must be made via external variables in a manner appropriate for randomized experimental trials of a given phenomenon. 

LA is an operationally motivated principle which captures the idea that any intervention could be made in a given round of an experiment---independently of any other variables that are causally relevant to the rest of the experiment, such as the initial states of the various systems under study---and that the intervention may only influence events in its future light cone. AOE is the assumption that any observed event has an absolute value, rather than relative to the perspective of an observer or to a branch of the wavefunction. It is valid, for example, in classical relativistic mechanics, where the same physical events occur from the reference frame of different observers. AOE is also implicitly assumed \cite{Bong2020, Wiseman2017} in the context of ordinary Bell scenarios, where the observed events refer to the outcomes that are recorded each round.

In an ordinary bipartite Bell scenario, the conjunction of these two premises allows the behaviour to be anything compatible with a general no-signalling model, as the assumption of AOE does not require the existence of any `elements of reality' or local hidden variables predetermining the outcomes of unperformed experiments. While Local Friendliness may seem innocuous from this perspective, it has nontrivial consequences in the context of extended Wigner's friend scenarios.

In EWF scenarios like the one described in Sec. \ref{EWFS} , the assumption of AOE effectively imposes that the outcomes observed by the friends possess values that have an existence independently of whether a superobserver decides  to read that value or not. This allows one to define a joint distribution $P(A,B,C,D|X,Y)$ over all the variables that are observed by some observer in every run of the experiment in the EWFS depicted in Fig.~\ref{fig:spacetime}.

The set of correlations $\wp(A,B|X,Y)$ for the superobservers consistent with the LF hypothesis in such scenarios has been shown to form a convex polytope, and so can be characterized by a finite set of inequalities. The LF polytope is in general a subset of the no-signaling polytope and always contains the corresponding Bell polytope, with example scenarios where the latter inclusion is strict highlighting the fact that the constraints from Local Friendliness are in general weaker than the hypothesis of LHVs \cite{Bong2020}. 

\section{A no-go theorem for Possibilistic Local Friendliness}

From the perspective of seeking an inequality-free no-go theorem, Def.~\ref{LA} needs to be revised, as it is phrased in terms of probabilistic notions. Given a specification of an experimental context, a physical theory may put limits to what types of behaviour can be observed in a given run of the experiment. From the perspective of a probabilistic theory, such as operational quantum theory, multiple outcomes could be considered possible. If the probability for an event to occur, within a given experimental context, is predicted to be zero, we say that this event is (physically) impossible in this experimental context. We propose the following possibilistic locality condition, which turns out to be sufficient for our purpose. 
\begin{definition}[Possibilistic Local Agency (PLA)] \label{DefPLA}
If a set of relevant events $E$ is possible, then for any intervention $Z$ not in the past light cone of any of the events in $E$, $E$ is possible in conjuction with any value of $Z$.
\end{definition}

Intuitively, PLA is the assumption that an intervention cannot influence possibilities of events outside its future light cone. While this way of expressing PLA would display its connection to LA more clearly, we prefer the phrasing in Def.~\ref{DefPLA} as it leaves no ambiguity about its logical implications.

LA implies PLA, but the converse does not hold. PLA is evidently guaranteed when considering the possibilities arising from probabilistic models that satisfy LA. This is because by the definition of LA, an intervention cannot affect the probabilities of events not in its FLC and hence the possibilities of such events are also unchanged. On the other hand it is easy to see that some probabilistic models violating LA obey PLA, e.g.~a model that allows for signalling but where no events are impossible.

We call the conjunction of AOE and PLA \emph{Possibilistic Local Friendliness} (PLF). For the reasons above, PLF is a weaker assumption than LF. 

For superobservers in the type of spatiotemporal relations of Fig.~\ref{fig:spacetime}, PLF implies that the possibilities in Alice's lab are unchanged by Bob's intervention and vice versa, which we call \emph{Possibilistic No-Signalling} (PNS) (a similar assumption was defined in refs.~\cite{Abramsky2013, Mansfield2012}). The fact that the converse does not generally hold is the content of the following theorem.

\begin{theorem} \label{thm:PLF_general} There exists Possibilistic No-Signalling phenomena in an EWFS which violate Possibilistic Local Friendliness.
 
\end{theorem}

\emph{Sketch of the proof.|}
We will use the two-party two-friend EWFS described before to demonstrate that the assumptions of AOE and PLA cannot in general be maintained together even if the possibilities for the superobservers obey PNS.

Suppose that the conditions of the EWF experiment are such that the events
\begin{align}
E_1 =^{def} (A=1,B=1,X=2,Y=2)\label{e1}
\end{align} are considered to be possible together, while the events
\begin{align}
 & E_2 =^{def}   (A=0,B=1,X=1,Y=2)\label{e2} \\
  &  E_3 =^{def}  (A=1,B=0,X=2,Y=1)\label{e3} \\
 & E_4 =^{def}  (A=1,B=1,X=1,Y=1) \label{e4}
\end{align} are considered to be impossible for Alice and Bob in the spatio-temporal configuration displayed in Fig.~\ref{fig:spacetime}. For example, a probabilistic theory could give $E_1$ a nonzero probability, while for $E_2$\,-\,$E_4$ a zero probability. These conditions are consistent with the principle of PNS, for example, if all the other unspecified sets of events are possible.

Now by the assumption of AOE, the event $E_1$ can be thought to occur along with some well-defined values of the outcomes $c,d$. So that in fact the set of events
\begin{align}
    (A=1,B=1,C=c,D=d,X=2,Y=2)\label{possible1}
\end{align}
can be considered possible, for at least some $c,d$, while the sets $E_2-E_4$ ought to be impossible in conjunction with any values for $c,d$. 

Next, we consider interventions  that would change the value of $X$, $Y$, or both. According to PLA, such interventions cannot affect the possibilities of events outside their respective FLC. We thus conclude from (\ref{possible1}) that the following sets of events are possible at least for some $a,a',b,b'$:
\begin{align}
    &(A=a,B=1,C=c,D=d,X=1,Y=2) \label{PLApossible1} \\
    &(A=1,B=b, C=c, D=d, X=2, Y=1)\label{PLApossible2}\\
    &(A=a',B=b',C=c,D=d,X=1,Y=1)\label{PLApossible3}\,.
\end{align}
 Note that the values of $c,d$ are the same in each of the lines (\ref{possible1}-\ref{PLApossible3}), as neither of the interventions are in their past light cones.  By construction of the protocol, along with AOE, when $X=1$, $A=c$ and when $Y=1$, $B=d$.  These imply that the following sets are possible:
\begin{align}
    &(A=c,B=1,C=c,D=d,X=1,Y=2) \label{PLAdoorpossible1}\\ 
    &(A=1,B=d, C=c, D=d, X=2, Y=1)\label{PLAdoorpossible2}\\
    &(A=c,B=d,C=c,D=d,X=1,Y=1)\label{PLAdoorpossible3}\,.
\end{align}

Then by line (\ref{PLAdoorpossible1}) and consistency with the impossibility of $E_2$ it must be that $c\neq0$. Similarly by line (\ref{PLAdoorpossible2}) and consistency with $E_3$ it must be that $d\neq0$. Finally by line (\ref{PLAdoorpossible3}) with $E_4$ $c,d \neq 1$. This is a contradiction, since now all the potential values for $c,d$ are exhausted. Hence the assumptions of AOE and PLA cannot be maintained together with the possibilistic conditions laid out on lines (\ref{e1}-\ref{e4}) in an EWFS.

Notably the possibilistic conditions on lines (\ref{e1}-\ref{e4})  are exactly of the kind that lead to Hardy's paradox \cite{Hardy1992, Hardy1993} and can be reproduced in quantum mechanics. For completeness, it will be shown in the latter part of the paper that such phenomena can be lifted to the EWF scenario.

It should perhaps at this point be emphasized that, despite using a similar possibilistic structure,  the no-go result sketched here genuinely differs from that of Hardy \cite{Hardy1992, Hardy1993}, which is exhibited in an ordinary Bell scenario. No contradiction  with PLF can emerge in ordinary Bell scenarios if Alice and Bob obey PNS, since in a Bell scenario PLF imposes no further constraint than PNS. This illustrates that the assumptions of AOE and PLA are strictly weaker than the assumptions that underlie Hardy's paradox. In order to get a no-go theorem in an ordinary Bell scenario, AOE and PLA need to be supplemented with additional assumptions, such as Predetermination~\cite{Wiseman2017}, which in our language would correspond to the requirement that for every set of interventions there is exactly one set of possible outcomes.  This is in analogy with the original LF no-go result, where the assumptions of AOE and LA if supplemented with the assumption of Predetermination lead to Bell inequalities \cite{Cavalcanti2021}.  In this sense, similarly to the LF no-go result, PLF on its own only becomes a nontrivial requirement in EWF scenarios.

We will next formalise the line of reasoning presented here using the Kripke-semantics of modal logic.

\section{Basics of modal logic}

The brief overview in this section is mostly based on ref \cite{Rantala2004}. The alphabet for the modal logic we consider consists of a set $\Omega$ of propositional variables $p_1, p_2 \ldots\in \Omega$, the familiar logical connectives $\neg, \wedge, \vee, \leftrightarrow, \rightarrow$  together with the modal operators $\diamondsuit$ for ``possibly" and  $\square$ for ``necessarily" along with the parentheses ( ). More complex formulas are formed from these by the following conditions: 
\begin{align*}
&\textbf{C1: } \textrm{All the propositional variables are formulas}.\\
&\textbf{C2: } \textrm{If $Q$ is a formula, then $\neg Q, \diamondsuit Q$ and $\square Q $ are formulas}.\\
&\textbf{C3: } \textrm{If $Q$ and $G$ are formulas, then so are } (Q\wedge G), \\ 
&\hspace{0,8cm}  (Q  \vee G,), (Q \leftrightarrow G) \textrm{ and } (Q \rightarrow G).\\
& \textbf{C4: } \textrm{There are no other formulas.}
\end{align*} Following the usual convention, the parentheses may be dropped for readability where there is no risk of confusion.

The common semantics for many modal logics is based on the so called Kripke-models. A Kripke-model $M$ is a triple $M = \langle W, R, V \rangle$, where $W$ is a non-empty set of \emph{possible worlds}, $R  \subset W \times W$ is a binary relation between worlds called \emph{accessibility relation} and $V$ is a $\emph{valuation function}$, i.e.~a mapping $V: \Omega \rightarrow \mathcal{P}(W) $ from the set of propositional variables  to subsets of $W$, indicating in which worlds the propositional variables are true. The notation $wRw'$ is used to indicate that $w'$ is accessible from $w$, or that $w'$ is a possible world from the perspective of $w$.

The notation $M,w \vDash Q$ (`$\vDash$' is read as ``entails'')  is used when $Q$ is  true in the world $w$ of the model $M$, and $ M,w \nvDash Q$ when it is untrue. For a $p \in \Omega$, $M,w \vDash p$ is equivalent to saying $w \in V(p)$. The truth conditions for the propositional connectives are defined as in ordinary propositional logic by the conditions
\begin{align}
& M,w \vDash \neg Q  \textrm{ iff } M, w \nvDash Q \label{truthnot},\\
&M,w \vDash G \wedge Q  \textrm{ iff } M,w \vDash G \textrm{ and } M,w \vDash Q \label{truthand},\\
& M,w \vDash G \vee Q \textrm{ iff } M, w \vDash G \textrm{ or } M,w \vDash Q \label{truthor},\\
&M,w \vDash G \rightarrow Q \textrm{ iff }  M,w \nvDash G \textrm{ or } M,w \vDash Q, \label{truthimplication}
\end{align}
and so on. Which is to say, that the truth conditions of classical propositional logic hold in each individual world. Here ``iff'' is short for ``if and only if".
 
 The truth conditions for the modal formulas are defined as follows:
 \begin{align}
 &M,w \vDash \square Q \textrm{ iff } M,w' \vDash Q \hspace{0,2cm}\forall w' \textrm{ s.t. } wRw'  \label{truthsquare}\\
 &M,w  \vDash \diamondsuit Q \textrm{ iff } M,w' \vDash Q \textrm{ for some } w' \textrm{ s.t. } wRw' \label{truthdiamond}
 \end{align}
 If $ \forall w \in W$, it holds that $M,w \vDash Q $, then  the formula $Q$ is said to be valid in the model $M$. The notation $M \vDash Q$ is used when $Q$ is valid in the model.
 
Notably, when evaluating modal statements like $\square Q$, one only needs to account for the worlds that are accessible from the reference world $w$.
The accessibility relation plays a very important role along the set of possible worlds $W$. For example, the intuitive statement $(\square Q \rightarrow Q)$, meaning ``if necessarily $Q$ then $Q$"  may not be valid, unless $wRw$ ie. the relation is reflexive. Another intuitive statement $\square Q \rightarrow \diamondsuit Q$ can fail unless every world has access to some world. We will not enforce any general conditions on $R$ based on  intuitive readings of $\square$ or $\diamondsuit$ and instead aim to phrase the argument  based on minimal assumptions.

\section{Formal proof of theorem 4}Let $M = \langle W,R,V \rangle$ be a model defined relative to a set $\Omega$ of propositional variables, the elements of which are interpreted to represent statements about physical observations in an EWFS of the type $(A=0), (B=0)...$ and so on. 
The notation  $(A=0,B=0\ldots Y=1)$ is used for conditions of the type $(A=0) \wedge (B=0) \wedge \ldots \wedge (Y=1)$.

Let $w_0$ be a world of reference where the possibilities related to the EWFS are evaluated. The assumptions about the possibilities in the experiment on lines (\ref{e1}-\ref{e4}) directly translate to
\begin{align}
    M,w_0 \vDash \diamondsuit E_1  \wedge \neg \diamondsuit  E_2  \wedge \neg \diamondsuit  E_3  \wedge \neg \diamondsuit E_4 \label{e1-e4},
\end{align} 
By AOE, each variable must have exactly one value in every possible world. That is, for every  $F\in \{A,B,C,D\},$
\begin{align}
 M, w_0 \vDash \square ((F=0) \wedge \neg(F=1)) \vee (\neg (F=0)\wedge (F=1)), \label{resultsmust}
\end{align}  
with similar conditions for $X,Y$.
PLA implies that, for any set of events $E$, and for any value $z$ of an intervention $Z$ not in the past light cone of any of the events in $E$,
\begin{align}
     M,w_0 \vDash (\diamondsuit E  \rightarrow \diamondsuit (E,  Z=z)). \label{PLA}
\end{align}

 Finally, the protocol for measurement choices $X=1$ and $Y=1$ corresponds to 
    \begin{align}
        &M,w_0 \vDash \square ((X=1) \rightarrow (A=C)) \label{Alicedoor}\\
       &M,w_0 \vDash \square( (Y=1) \rightarrow (B=D)) \label{Bobdoor},
    \end{align}  where $A=C$ is shorthand for 
\begin{align}
    ((A=0)\wedge (C=0)) \vee ((A=1)\wedge (C=1)).
\end{align}
Technically, for consistency with the physical picture, the $\vee$ above should be an exclusive-or, but this detail makes no difference for the argument and in any case, that requirement is guaranteed by (\ref{resultsmust}).

Now by (\ref{truthand}) the assumption (\ref{e1-e4}) can be true only if:
\begin{align}
    M,w_0 \vDash \diamondsuit E_1 \label{diamondp1}
\end{align}
\begin{figure}
    \centering
    \includegraphics[scale=0.3]{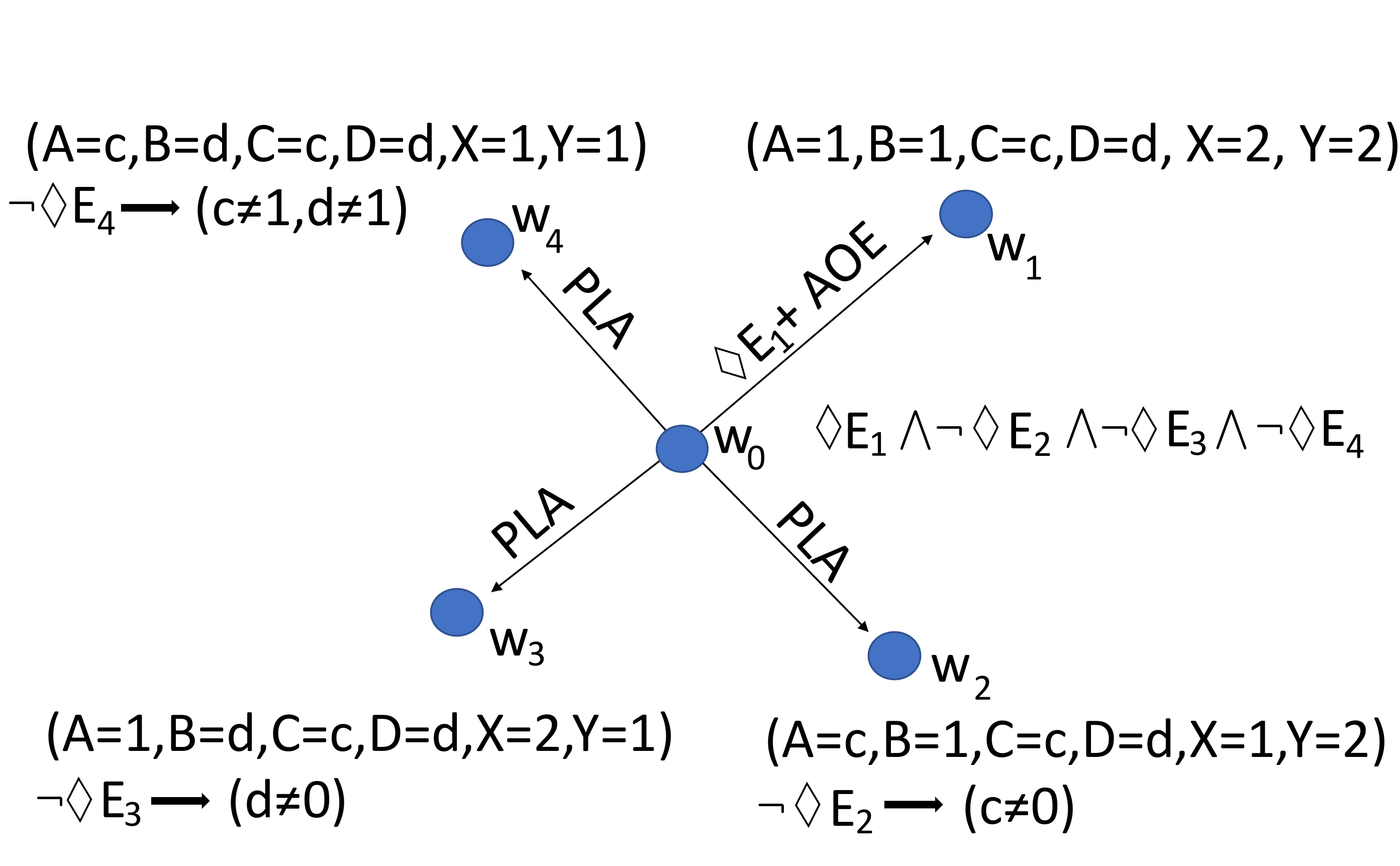}
    \caption{Graphical representation of the argument. $\diamondsuit E_1$ and AOE imply the existence of an accessible world $w_1$ in which $(E_1,C=c,D=d)$ is true for some values of $C,D$. PLA with the space-time relations of the variables then imply the existence of three accessible worlds $w_2,w_3,w_4$ in which different interventions $X,Y$ have been chosen, with the same fixed values $C,D$. Once the reading of $X,Y =1$ is taken into account, it follows that if $C=0$ then $w_2$ is inconsistent with $\neg \diamondsuit E_2$, if $D=0$ then $w_3$ conflicts with $\neg \diamondsuit E_3$  and $w_4$ conflicts with $\neg \diamondsuit E_4$ when $C=1,D=1$ so that  a contradiction cannot be avoided.}
     \label{fig:worldnetwork}
\end{figure}

The truth conditions for  $\square, \wedge, \vee$ allow one to combine the use of line (\ref{resultsmust}) for $C,D$ with line (\ref{diamondp1}) to get the condition
   \begin{align}
     \exists (c,d)\,  &M,w_0 \vDash \diamondsuit (E_1\wedge (C=c)\wedge (D=d)) \label{possibleCD},
   \end{align}
   where the terms with negations have been dropped from formulas of the type $(F=f)\wedge \neg(F=f')$ for $C,D$  by using the implication $\diamondsuit (Q\wedge G) \rightarrow \diamondsuit Q$ which follows from the truth conditions for $\wedge$. Here, the `$\exists$' over $c,d$ is understood as a disjunction of propositions of the above type over all possible values of $c,d$.
   
 From (\ref{possibleCD}) onwards, the proof follows the earlier sketch and will not be repeated in full. The crux of the argument is that for any $c,d$ one can use PLA and combine with (\ref{resultsmust}, \ref{Alicedoor}, \ref{Bobdoor}) in a way that implies a contradiction with (\ref{e1-e4}). This thought chain is depicted compactly in terms of the network structure of the Kripke model in Fig.~\ref{fig:worldnetwork}.

The argument presented here is independent of quantum mechanics. Quantum theory does, however, in principle allow for the type of possibilistic conditions described on (\ref{e1-e4}) and hence, any theory compatible with quantum predictions leads to similar contradictions and therefore has to abandon AOE or PLA. This is the content of theorem \ref{theorem} below.

    \begin{theorem} There exist quantum predictions on an EWFS that are incompatible with Possibilistic Local Friendliness. \label{theorem}
     \end{theorem}
    \proof
   
    Consider the quantum-mechanical model in which Charlie and Debbie share a qubit pair prepared in state
\begin{align}
    \ket{\psi} = \dfrac{1}{\sqrt{3}} ( \ket{0}_{S_A}  \ket{0}_{S_B} + \ket{0}_{S_A} \ket{1}_{S_B} + \ket{1}_{S_A} \ket{0}_{S_B}) ,
\end{align} 
where $\ket{0}_{S_{A/B}}, \ket{1}_{S_{A/B}}$ form an orthonormal basis for $\mathbb{C}^2 \simeq \mathcal{H}_{S_A}, \mathcal{H}_{S_B}$.  Charlie and  Debbie perform two-outcome measurements which are described, from their perspective, with projective elements $\ket{0}\bra{0}_{S_{A/B}}, \ket{1}\bra{1}_{S_{A/B}}$. From the perspective of Alice and Bob, the laboratories including $S_A$ and $S_B$ evolve unitarily to 
\begin{align}
    \ket{\Psi} &= U_A \otimes U_B [\ket{R}_{F_A} \otimes \ket{\psi}_{S_A S_B} \otimes \ket{R}_{F_B}]\nonumber\\
    &=  \dfrac{1}{\sqrt{3}} (\ket{C_0} \otimes \ket{D_0} + \ket{C_0} \otimes \ket{D_1} + \ket{C_1} \otimes \ket{D_0}) \textrm.
\end{align}  
Here $\ket{R}_{F_A} \in \mathcal{H}_{F_A}$   and  $\ket{R}_{F_B} \in \mathcal{H}_{F_B}$   refer to the ``ready'' state vectors of Charlie and Debbie and their laboratories in the beginning of the experiment; $U_A$ and $U_B$ are the unitaries that correlate each laboratory with their qubit upon measurement; $\ket{C_0}, \ket{C_1}$ refer to orthogonal state vectors of the joint system $F_A$ and $S_A$ after Charlie having observed outcome $0 /1$, and similarly for Debbie.

 Alice's measurement choice corresponding to the setting $X = 1$ may consist of the elements $A_1(a) \in  \{ \ket{C_0}\bra{C_0}, I_{S_A,F_A} - \ket{C_0}\bra{C_0} \}$ and similarly for Bob $B_1(b) \in \{ \ket{D_0}\bra{D_0}, I_{S_B,FB} - \ket{D_0}\bra{D_0} \}$. Let the settings $X=2 $ and $Y=2$ correspond to the projective measurements 
\begin{align}
\begin{matrix}
A_2(0) = \ket{C_+} \bra{C_+}  & A_2(1) = I_{S_A,F_A} - \ket{C_+} \bra{C_+}\\
B_2(0) = \ket{D_+} \bra{D_+} & \hspace{0,2cm}B_2(1) = I_{S_B,F_B}- \ket{D_+}\bra{D_+},
\end{matrix} 
\end{align}
where $\ket{C_{+}} = \dfrac{1}{\sqrt{2}}[\ket{C_0} + \ket{C_1}]$ and so on. The proof concludes by noting that a direct calculation of $P(ab |x y) = \bra{\Psi} A_{x}(a) \otimes B_{y}(b) \ket{\Psi}$ produces the possibilistic structure used in the proof of Theorem~\ref{thm:PLF_general}: 
\begin{align*}
 P(1,1|1,1) &= 0, P(0,1|1,2) = 0 \\ P(1,0|2,1) &= 0 , P(1,1 | 2,2) > 0 \textrm.
\end{align*} 
\section{Conclusions} We have constructed a possibilistic version of the Local Friendliness no-go theorem by exhibiting a logical contradiction under the assumption that the possibilities for the superobservers in an Extended Wigner's Friend Scenario are of the type that lead to Hardy's paradox. The result does not presuppose quantum mechanics. However, under the assumption that quantum theory holds even on the scale of observers, and that some such observers (e.g.~AI programs in a future large quantum computer) could be in principle coherently controlled by superobservers, one gets a no-go theorem using the type of possibilities allowed by quantum theory.  The conclusion is of foundational interest, as the assumptions that lead to this Possibilistic LF no-go theorem are weaker than those of the original LF no-go theorem~\cite{Bong2020}, which are in turn weaker than those required to derive Bell's theorem.

The Possibilistic LF theorem refers to idealised predictions, and is thus not amenable to experimental test---similarly to the Hardy and GHZ paradoxes, and unlike Bell's theorem and the original LF no-go theorem. Nevertheless, it shows that any resolution of the contradiction in the original LF no-go theorem that merely involves the use of probability theory in that proof cannot apply in the ideal limit. It thus provides a logical argument against that alternative.

While we have demonstrated our result by considering a very particular EWFS, the definition of PLA and the formal approach using the Kripke semantics of modal logic should be readily adaptable to the consideration of more general scenarios, with different numbers of superobservers and friends respectively. It would be interesting to see, for example, if some multipartite EWF scenarios would allow for demonstrations of a PLF no-go theorem with a possibilistic structure of the GHZ type \cite{Greenberger1989, Greenberger1990}.   More general questions of interest include the formulation of necessary or sufficient possibilistic conditions for contradictions in various EWFS and whether or not such conditions could be  reproducible in quantum theory. 
 
From another perspective, this result casts some doubt on whether or not a logic described with  Kripke-structures can be correct for reasoning about the possible courses of events in space-time allowed for by general theories that satisfy PLA. Indeed, the emergence of a contradiction suggests that a `complete story' of physically possible events in relativistic space-time shouldn't perhaps be told in terms of propositions with definite truth values for all observations. To repeat a point made in ref \cite{Nurgalieva2019}; it may be that the possible worlds in Kripke models are too classical for quantum (or more general) settings. In light of this, it would be interesting to see if the modal logic framework could be generalized somehow to be more consistent with the quantum phenomena, while reducing to the classical notions for events about which information is accessible to the superobservers at the end of the experiment. 

\section*{ACKNOWLEDGEMENTS}
This work was motivated by a discussion with Jacob Barandes and Matt Leifer following a talk presented by EGC~\cite{Harvard}. EGC acknowledges useful discussions with Cristhiano Duarte on the topic of modal logic. We thank Howard Wiseman and anonymous referees for the Quantum Physics and Logic 2022 conference for feedback on a draft of the paper. This work was supported by the Australian Research Council Future Fellowship FT180100317, and grant number FQXi-RFP-CPW-2019 from the Foundational Questions Institute and Fetzer Franklin Fund, a donor advised fund of Silicon Valley Community Foundation.

\bibliography{PossibilisticHref}
\end{document}